\documentstyle[aps,preprint,epsfig]{revtex}
\newcommand{\bat}{$\rm BaTiO_{3}$}
\newcommand{\knb}{$\rm KNbO_{3}$}
\newcommand{\dwo}{$180^\circ$}
\newcommand{\dwn}{$90^\circ$}
\newcommand{\dwfv}{$45^{\circ}$}
\begin{document}
\title
{Efficient Switching and Domain Interlocking\\
Observed in Polyaxial Ferroelectrics} \author {A.
Krishnan,$^1$ M.M.J. Treacy,$^1$ M.E. Bisher,$^1$ \\P. Chandra$^1$ and
P.B. Littlewood$^2$}
\address{$^1$NEC Research Institute, 4
Independence Way, Princeton NJ 08540\\$^2$Cavendish
Laboratory, Madingley Road, Cambridge CB3 OHE, UK}
\maketitle
\begin{abstract}
\setlength{\baselineskip}{18pt}
We present transmission electron microscopy observations of domain wall
motion in thin freestanding \knb\ crystals under applied electric
fields. Since there is no substrate, there is no elastic clamping of
\dwn\ domains. We observe that curved and tilted \dwn\ domain walls are
the most mobile, whereas untilted \dwn\ domain walls are resistant to 
field-induced motion.

We explain this result in terms of two factors. First, the switching
pressure on a domain wall $(\mbox{\boldmath $P$}_{2} \!  - \!
\mbox{\boldmath $P$}_{1})\!\cdot\!\mbox{\boldmath $E$}$ is determined by
the relative electrostatic energies of the neighboring polarizations
$\mbox{\boldmath $P$}_{1}$ and $\mbox{\boldmath $P$}_{2}$. Consequently,
some \dwn\ domain walls are immobile under certain field directions,
leading to domain interlocking. Second, domain walls experiencing a high
switching pressure move by a ripple mechanism, and do not move as rigid
sheets. The tilted wall region in such a ripple has a polarization
charge, and an associated depolarization field, which reduces the local
switching barrier. An accumulation of polarization charge can result in
a tilted or curved wall, as occurs at the mobile tips of \dwn\ domain
needles.

Any increase in density of immobile wall configurations with cycle time
represents an inherent contribution to fatigue.  Uniaxial
ferroelectrics, with polarizations parallel to the field, should not
experience such domain interlocking.
\end{abstract}

\pagebreak

\section{Introduction}
Ferroelectric materials are attracting much interest because of their
potential use as non-volatile computer memories \cite{Scott00}.
However, the physical processes that lead to degradation effects, such
as fatigue and imprint, are still incompletely understood and represent
significant barriers to the development of ferroelectric devices.
Ferroelectric fatigue appears as a gradual reduction of the switchable
polarization with time, whereas imprint occurs when one polarization
direction requires a larger coercive field than does the reverse
polarization.  It is clear that fatigue is due, in part, to the
suppressed motion of domain boundaries, which inhibits the polarization
reversal processes.  The link between the increased density of
unswitchable domains and fatigue has been established by direct
observation using atomic force microscopy \cite{Colla98a}. Several
contributions to the degradation of ferroelectric capacitors have been
identified, leading to the development of improved devices
\cite{Ramesh93,Nakamura94,DeAraujo95,Scott98,Damjanovic98}; these
include charge accumulation near the electrodes \cite{Colla98}, and
electromigration of extrinsic pinning centers such as charged defects
\cite{Warren95a} and oxygen vacancies \cite{Scott89}.

In this paper, we present in-situ transmission electron microscope (TEM)
observations on single crystal \knb\ which indicate that domain wall
motion can be inhibited in homogeneous polyaxial ferroelectrics.  Our
studies are conducted at temperatures much less than the Curie
temperature in electric fields that are well below the nominal coercive
value.  In our experiments, the ferroelectrics are freestanding and are
not clamped by any support material. The observed domain configurations
are robust to mechanical tapping, and in the absence of an electric
field, are stable for long periods of time. No spontaneous nucleation is
observed, implying that switching is dominated by domain wall motion.

The switching pressure on a domain wall is given by the difference in
free energy of small regions either side of the wall. We observe that in
a polyaxial ferroelectric, where many different domain wall orientations
can occur, not all domains of a given polarization are equally
switchable under a fixed electric field. In particular, if the field
direction is reversed, the domain response changes. Although elastic
interactions are clearly important in the energetics of switching, this
observed behavior cannot be explained by piezoelectric interactions
alone, since at equilibrium the strain energy depends only on even
orders of the polarization and is not very sensitive to changes in wall
domain orientation and electric field direction. By contrast, the
electrostatic energy difference, $(\mbox{\boldmath $P$}_{2} \! - \!
\mbox{\boldmath $P$}_{1})\!\cdot\!\mbox{\boldmath $E$}$, across a wall
bordering domains of polarization $\mbox{\boldmath $P$}_{1}$ and
$\mbox{\boldmath $P$}_{2}$ is affected by the field direction.
Consequently, not all walls are equally mobile. Therefore, in a
polyaxial ferroelectric, not all domains of a given polarization are
equally switchable under an applied electric field.

In this paper we explain our results phenomenologically in terms of the
electrostatic energy contributions. Although elastic energies are
clearly important, they alone cannot explain the observed
neighbor-dependent switching efficiency of domains. This implies an
inherent distribution of coercive fields for domain switching. Our
results suggest that polyaxial ferroelectrics are vulnerable to any
redistribution of domain size and configuration with cycle time that
leads to an increase in density of the most inhibited domain
configurations.  We discuss the implications for fatigue and imprint in
polyaxial ferroelectrics.

\section{Experimental Results}
In-situ TEM studies of ferroelectrics provide an opportunity to observe
extended areas of domain structure at high resolution while
simultaneously modifying external parameters
\cite{Tanaka64,Hu86,Yamamoto80,Snoeck94}.  Here we report results from
observations of freestanding single-crystal potassium niobate \knb,
noting that we have seen similar phenomena in single-crystal barium
titanate \bat. All experiments were performed in a Hitachi H9000-NAR TEM
at 300 kV at room temperature, well below the Curie temperature $T_{\rm
c}$.  The samples were mounted in a custom-modified Gatan single-tilt
holder that allowed application of electric fields to the sample during
observation. In the absence of electric fields, domain configurations
are stable to mechanical handling. Experimental details have been
reported elsewhere \cite{Krishnan98}.

Figure 1 shows micrographs of an area of thin \knb, viewed along one of
the principal pseudo-cubic $\left< 100 \right>$ axes, before and after
application of an electric field. Prior to acquiring this pair of
micrographs, the specimen was subjected to approximately ten field
cyclings, and thus Fig.\ 1a does not represent a virgin state.  The
in-plane direction of the electric field, as determined by the electrode
geometry, is indicated. There is also a vertical (i.e.\ into the page)
component of the field. The amplitude of the field in this region is
estimated to be $<0.2$ V.$\mu$m$^{-1}$.  As frequently happens in these
domain switching experiments, the region of interest moved rapidly out
of the field of view when the electric field was applied.  Consequently,
this leaves us with just the before and after views. Those domain walls
that are aligned edge-on to the viewing direction show up as dark thin
lines primarily because of diffraction contrast caused by shear strain
across the boundary.  Although \knb\ can support many types of domain
wall structure, the area shown in Fig.\ 1 is dominated by \dwn\ and
\dwo\ domain walls, which are the most common.  Initially (Fig.\ 1a) we
observe many needle-like \dwn\ domain walls that are pinned at their
tips by a third domain orientation; by \dwn\ and \dwo\ we refer to the
approximate relative angle between polarization vectors,
$\mbox{\boldmath $P$}_{1}$ and $\mbox{\boldmath $P$}_{2}$, adjacent to a
given boundary.  The polarization charge surface density, $\sigma \!  =
\! \mbox{\boldmath $\Delta P$}_{\perp} \!  = \!  (\mbox{\boldmath
$P$}_{1} \!  - \!  \mbox{\boldmath $P$}_{2})\!\cdot\!\mbox{\boldmath
$\hat{n}$}$, associated with a domain wall is zero if the normal
component of the polarization, $\mbox{\boldmath $P$} \!\cdot\!
\mbox{\boldmath $\hat{n}$}$, is continuous across the interface. Here,
{\boldmath $\hat{n}$} is the unit vector normal to the wall into domain
$\mbox{\boldmath $P$}_{2}$.  Curvature in domain boundaries inevitably
implies discontinuities in $\mbox{\boldmath $P$}\!\cdot\!\mbox{\boldmath
$\hat{n}$}$ across parts of the boundary, resulting in non-uniformly
charged walls and the possibility of increased mobility under an applied
electric field, $\mbox{\boldmath $E$}$.  We note that in both \knb\ and
\bat\ the observed domain patterns are stable to mechanical
perturbations and to thermal cycling within the ferroelectric phase
\cite{Krishnan98}, indicating that ferroelastic effects do not dominate
the observed domain behavior. The weak, but observable, contrast at both
the \dwn\ and charge-neutral \dwo\ domain walls is confirmed by standard
multislice TEM image simulations \cite{Kilaas88}.

In Figs.\ 1c(d) we delineate the relevant domain boundaries observed in
the micrographs, along with the associated polarization directions and
boundary charges.  Selected area diffraction confirms the orientation of
the polarization axes to within an ambiguity of $\pm 180^{\circ}$.  For
consistency with the known electric field direction, we associate
positive charge with the curved domain boundaries $A_{1}$, $A_{2}$,
$C_{1}$, $C_{2}$, $C_{3}$ whereas the straight interfaces along the
diagonal wall are neutral.  This interpretation is confirmed by
temporary application of a weak electric field, as displayed in Figs.\
1b and 1d (the field is switched off in Fig.\ 1d).  Here we observe that
curved boundaries and associated tips have moved, whereas the neutral
\dwo\ and \dwn\ interfaces (e.g. $B_1$) have undergone little change,
suggesting that an applied electric field has maximum effect on curved
and tilted walls.  We have micrographs of qualitatively similar
field-induced domain behavior in \bat, although the details of needle
shape and relative abundance of \dwn\ and \dwo\ walls are different.

\section{Discussion}

Domain walls can minimize electrostatic energy by maintaining continuity
of the normal component of polarization $\mbox{\boldmath $P$} \!\cdot\!
\hat{\mbox{\boldmath $n$}}$ across the wall.  This condition ensures
zero polarization charge density on the wall and consequently no local
contribution to the depolarization field. However, if the domain wall
were tilted by an angle $\theta$ relative to the ideal angle (see Figure
2) then a charge density $\sigma$ is developed.  Ignoring small
corrections due to deviations from true cubic symmetry, the surface
charge density at \dwn\ domains (Fig.\ 2a) is $\sigma_{90} =
\sqrt{2}P\sin\theta$, and at \dwo\ domains (Fig.\ 2b) is $\sigma_{180} =
2P\sin\theta$.

We observe experimentally in \knb\ and \bat\ that domain wall curvature
and tilt often occurs when perpendicular sets of \dwn\ domains
intersect.  Such tip curvature can result from a competition between
minimization of the electrostatic energy ($U_{E} \propto \int \sigma^2
dA$) and conservation of polarization charge ($Q \sim \int \sigma dA$)
at the interrupted domain tip.  We illustrate this point in Figure 3 for
the simplest case of planar polarization.  Figure 3a shows a sketch of
two $90^\circ$ domain orientations $A$ and $B$ where a third, $C$, is
contained in $B$.  The interface between $A$ and $C$ is a $180^\circ$
tilted ($\theta \approx 45^{\circ}$) wall with polarization charge
density $\sigma \approx \sqrt{2} P$ per unit area. Naturally this charge
density can be eliminated if all of $C$ transforms into $B$.  The
positive polarization charge then will flow to the opposite tip of
domain $C$ (or, conversely, negative polarization charge could flow from
the opposite tip, with the same end result), and the polarization
component $B$ increases.  The electrostatic energy of the configuration
in Fig.\ 3a can be lowered, while preserving approximately the total
polarization, by curving the interfaces enclosing $C$.  The resulting
non-uniform charge density reduces $U_{E}$ and involves a redistribution
of the boundaries between $B$ and $C$.  Alternatively, the wall between
$A$ and $C$ can also tilt through $45^{\circ}$ (Fig.\ 3c), resulting in
a zero polarization charge density, $\sigma \!  = \!  \mbox{\boldmath
$\Delta P$}_{\perp} \!  = \!  (\mbox{\boldmath $P$}_{1} \!  - \!
\mbox{\boldmath $P$}_{2})\!\cdot\!\mbox{\boldmath $\hat{n}$}=0$,
everywhere.  To accomplish such domain wall restructuring, with zero
charge at the interfaces, a negative polarization charge must flow in
from the left and right boundaries.  The development of this
charge-neutral pattern requires motion of the boundaries between $A$ and
$B$.  The relative importance of these configurations depends on the
energetic barriers between the different polarization domains $A$, $B$
and $C$, and here elastic effects could play an important role.

In multicomponent ferroelectrics, the spontaneous polarization is
accompanied by a local deformation which, for simplicity, we do not
treat directly in the present discussion.  Similar needle-like domain
structures to those discussed here (see Figure 3b) also arise in
ferroelastics \cite{Salje90} and martensitics \cite{Nishiyama78} due to
competition between topological and long-range elastic effects.  In the
context of ferroelectrics we emphasize the polarization charge
associated with such configurations, since this charge is crucial
towards understanding their field-induced response.

Under an applied electric field {\boldmath $E$}, domains switch so as to
reduce the polarization energy, $U_{P} = -\mbox{\boldmath $P$} \!\cdot\!
\mbox{\boldmath $E$}$.  However, at temperatures well below the Curie
temperature, $T_{\rm c}$, there is a barrier to switching. Above the
coercive field, this barrier disappears and switching occurs
effortlessly.  Our in-situ data, however, is acquired at fields well
below the theoretical coercive field, and indicates that switching is
dominated by domain wall motion rather than by spontaneous domain
creation. The switching pressure on the wall separating two domains with
polarizations $\mbox{\boldmath $P$}_{1}$ and $\mbox{\boldmath $P$}_{2}$
is given by $(\mbox{\boldmath $P$}_{2} \! - \! \mbox{\boldmath
$P$}_{1})\!\cdot\!\mbox{\boldmath $E$}$. This pressure is dependent on
field direction, and can be zero. In the micrographs of Figure 1, the 
fact that the domain walls at A$_{1}$, A$_{2}$ etc. are curved and 
tilted indicates that those parts of the wall are moving. These walls 
therefore, must be experiencing higher switching pressures than the 
straight walls, suchas those at B$_{1}$.

In the appendix, we present a simple isotropic Landau-Ginsburg Free
energy analysis of a tilted \dwo\ domain wall that is subjected to
applied field. The analysis confirms that there are two effects caused
by wall tilt, both of which increase the wall mobility under an applied
electric field.  First, the polarization energy term $\mbox{\boldmath
$P$} \!\cdot\! \mbox{\boldmath $E$}$ is enhanced by a factor $1
+g(\theta)$.  This enhancement counteracts the reduction in local
{\boldmath $P$} caused by the depolarization field. Consequently, the
pressure term $\left(1 + g(\theta)\right)\mbox{\boldmath $P$} \!\cdot\!
\mbox{\boldmath $E$}$ is approximately independent of angle and roughly
equal to $\mbox{\boldmath $P$}_{0} \!\cdot\! \mbox{\boldmath $E$}$ at
all wall tilt angles $\theta$, as expected. The second is that the
effective Curie temperature is lowered, which means that the switching
barrier height $\Delta F_{\rm barrier}$, which is of the order of
$k_{\rm B}\tilde{T}_{\rm c}(\theta)$, is lowered.  Untilted,
charge-neutral, walls have the highest barriers to switching.
Realistically, depolarization fields will be screened at large
length-scales, particularly in thin films, but will continue to
influence the polarizations adjacent to the wall under study.

The preceding free energy argument provides insight into the
field-induced motion of needle-like domains such as those shown in Fig.\
1b(d).  The curved tip of the needle supports a polarization charge that
increases with tilt and is maximum at the horizontal wall of the tip
(where $\theta = 90^{\circ}$, with $\theta_{1} = 0^{\circ}$).  In the
absence of an electric field, the switching barrier at the tip is
reduced compared with the uncharged needle sides.  Consequently, the
effective coercive field at the tip is lower than that of the flat
sides.  Under an applied field, the tip moves first, since the
electrostatic pressure is approximately constant over the wall, and the
barrier is lowest at the polarization-charged tip. It is believed that
domain walls are unstable under finite electrostatic pressure
\cite{Chervonobrodov88}, and ripples can appear. In order for the needle
to retract, the tip must transfer its charge to lower parts of the wall.
Most likely, the tip achieves this charge transfer by emitting a series
of charged ripples which flow along the wall \cite{Callaby65}.  These
ripples can propagate because the local charge at the ripple increases
the local wall angle, concomitantly reducing the local switching barrier
and increasing the local switching pressure. Locally, after the ripple
has passed, the wall has moved sideways thus increasing the volume of
polarization energetically favored by the applied field.  From another
perspective, these charge ripples result from local nucleation events at
domain walls, and a number of experiments suggest that lateral domain
motion is controlled by these processes
\cite{Jona93,Landauer57,Miller58}. In particular, the measured
exponential field-dependence of the wall velocity gives strong support
to this scenario \cite{Miller58}.

We now apply this understanding to the data shown in Figure 1.  Figure 4
illustrates schematically an interlocked domain system in a similar
orientation to that observed in Fig.\ 1.  Three out of the four possible
in-plane polarization directions are present.  Domains A, B and C have
the same polarization, with A and B forming vertical needles bordering
\dwn\ domains D, E and F, whereas C forms a horizontal needle bordered
by \dwn\ domains G and H, which have antiparallel polarization to D, E
and F. Domains antiparallel to A, B and C are not present.  In this
diagram the electric field moves positive charge from top to bottom, and
this flow is represented here as ripples of polarization charge $\Delta
Q_{\rm D}$ along the diagonal domain wall labeled 1--2--3--4--5.  Charge
flows from the upper left corner until it reaches the tip of domain A
(region 2 in Fig.\ 4a).  Domain A presents two additional domain wall
surfaces along which charge flow can branch.  Domain A cannot detach
until a minimum charge has been accumulated (Fig.\ 4b), which is
proportional to its width $W$.  In addition, the charge accumulating
near the tip 2 will generate a local depolarizing field which may act to
decrease the flow of polarization current along the diagonal wall.

Once domain A detaches, and its tip region 2 begins to retract (Fig.\
4c), it carries a fixed polarization charge.  The end of domain A
transforms into a pointed tip in order to spread the fixed polarization
charge $Q = \int \sigma dA$ over a larger surface area so as to reduce
the charge density, $\sigma$, and hence the electrostatic energy, $U_E
\propto \int \sigma^{2} dA$.  The charged ripples from the electrode now
continue along the diagonal wall, 1--3--4--5, to the tip of domain B
(region 3 in Fig.\ 4d).  The walls between domains B and G, and B and H,
can support positive charge ripples which would shrink B. However, there
is no driving force to propagate such ripples since the pressure on the
wall $(\mbox{\boldmath $P$}_{\rm G} \!  - \! \mbox{\boldmath $P$}_{\rm
B})\!\cdot\!\mbox{\boldmath $E$} \approx 0$. There is no energy
reduction obtained by switching B. Consequently, the charge flows past
B, moving its tip 3 by an amount equal to the ripple height, but leaving
the tip shape unaffected.  The charge ripple moves on to domain C
(Figs.\ 4e and 4f), which subsequently behaves similarly to domain A. In
the meantime, the tip of domain A has been retracting at a constant rate
due to electrostatic pressure from the electrodes.

Although A, B and C have equivalent polarization orientations, it is
clear that they are not equally mobile.  The curved tip of A can move
parallel to the field in order to reduce the total polarization energy
of the system, whereas the motion of B would be perpendicular to the
field, with no reduction in overall energy.  There is no switching force
on the wall separating B and G. The B/G and B/H walls are intrinsically
inhibited by the environment of domain B. B can only reduce its energy
by switching to the orientation of domain E at its tip, and that
switching rate is governed by the motion of the wall 1--2--3--4--5. Such
intrinsically domain-interlocked configurations can reduce the amount of
switchable polarization.  We note that nonlinear field effects can
induce boundary polarization charge \cite{Krishnan00}, supporting the
observation that periodic high-field pulses relieve ferroelectric
fatigue \cite{Scott88}.  We have observed that fine-scale domain
fingering patterns move faster than larger ones, consistent with the
notion of curvature-dependent charge densities.  Current theories of
ferroelectric switching times assume geometrically-independent domain
velocities \cite{Scott98}; the observations here clearly point to the
limitations of such approaches.

The sign and direction of the charge ripples in a particular domain
configuration are important.  Figure 5a depicts the same region as in
Figure 4a, under the same electric field, but this time the diagonal
wall is carrying negative charge in the opposite direction,
5--4--3--2--1.  Following a similar reasoning as that for Fig.\ 4, we
find that the principal difference in behavior is that in Figure 4,
domain A detaches first, whereas in Figure 5, domain C detaches first.
For different domain configurations and field orientations, the
asymmetry in field response could be quite large, and thus might be an
inherent contribution to imprint.

The illustrative example described above shares many similarities with
the domain wall behavior observed in Figure 1.  The domains $A_1$ and
$A_2$ are analogous to domain $A$ in Figure 4; similarly $B_1$ is
analogous to $B$ and $C_1$, $C_2$ and $C_3$ are analogous to $C$.  The
domains $A_1$ and $A_2$ have retracted the furthest under the field,
whereas $C_1$, $C_2$ and $C_3$ have traveled the least distance. $B_1$
is relatively unaffected by the field.  The position of the five tips is
consistent with $A_1$ having detached first, followed by $A_2$ etc.\ and
with subsequent uniform field-induced tip motion.  The switching rate of
$B_1$ is determined entirely by the charge flow along the diagonal wall
through its tip, and is therefore an indirect multi-step process.

Charge ripples play a loosely analogous role to that of dislocations in
the plastic deformation of crystals, where it is considerably easier to
propagate a single dislocation over a slip plane, than it is to shear a
whole region of the crystal over the same plane in one movement.  The
end result is the same, but the barrier to dislocation motion along the
slip plane is significantly lower and is therefore much more likely to
occur.

Our observations demonstrate that the switchability of an unsupported
ferroelectric depends on the distribution of domain orientations and
their widths.  More specifically, the charge threshold required for the
field-induced depinning of a domain is proportional to its width, so
domains of the type A and C in Figure 4 retract and hence switch more
readily than their wider counterparts.  By contrast, a domain of type B
in Figure 4, identical in polarization orientation to that of A and C,
requires multi-step switching processes since the polarization
orientations of its neighbors make direct switching energetically
unfavorable.  The switchability of a particular domain then depends not
only on the specifics of its own polarization orientation, but also on
its possible final states as defined by its adjacent polarizations.  We
also note that there is an anisotropy in the intrinsic switching
mechanism described here.  In Figure 4 if the field were applied
parallel to the diagonal wall, i.e.\ from the upper left to the lower
right, then the relative mobility of domains A, B and C changes.  Some
crystal orientations will be less vulnerable to inhibition of domain
switching.  This conclusion is consistent with recent experiments which
exhibit fatigue anisotropy in single-crystal ferroelectrics
\cite{Bornand00}.  It follows from our model that if the electric field
were to be rotated 90$^{\circ}$, the roles of mobile and jammed domains
would reverse.  If the field were applied from right to left in Figure
4, domain B would form a negatively charged mobile tip which would
retract to the right.  A, C, D, E and F are now the interlocked domains.
This suggests that the periodic application of a perpendicular field
would provide a method for flushing out domain-locked regions in a
fatigued polyaxial ferroelectric.  Earlier experiments on thin films of
fatigued lead zirconate titanate show that unfatigued ferroelectric
properties are indeed measured when the applied electric field is
changed by $90^{\circ}$ \cite{Pan92}.

It is well-known that the experimental coercive field is significantly
lower than the predicted theoretical value \cite{Jona93}.  As we have
shown, a tilted wall has a reduced coercive field compared to its
untilted counterpart because of the associated depolarization field. The
relevant angles to consider are not the static wall tilt angles observed
in our micrographs.  Instead, they are the angles adopted by the
propagating charge ripples.  We expect that these field-dependent angles
could be large, since the ability of the charged ripple to propagate
requires a highly reduced switching barrier.  We have not observed such
charge ripples directly in our experiments, since they move rapidly and
could be of small amplitude, and so we can not make a quantitative
analysis of this contribution. However, it is clear that domain walls do
not move as rigid structures, and that therefore polarization charges
are necessarily involved in domain wall motion.

Many models of ferroelectric fatigue involve the electromigration of
extrinsic pinning centers that suppress domain motion and hence
polarization reversal \cite{Robels95,Yoo92,Brennan93,Dawber00}.  In
zero-applied field we frequently observe domain walls entangled with
dislocations, such as shown in Figure 6.  Curved and tilted domain walls
can attract migrating charged defects electrostatically to neutralize
their surface charges.  Our studies indicate that these needle-like
domains are highly mobile in an applied field, and can move towards
extrinsic charge centers that are essentially stationary on the
appropriate short time-scale \cite{Pertsev88}.

The different domain behaviors we observe in our experiments are
definitely not elastic stress-related phenomena, since our samples are
freestanding single crystals. The electrodes are not in intimate contact
with the regions being observed.  Domain clamping by electrodes
\cite{Ganpule01}, although crucially important in epitaxially grown
ferroelectric systems, does not occur in our experiments. Stresses are
indeed present in our samples. The tip structures labeled 2, 3 and 4 in
Figure 4a, are similar for both mobile and immobile needles, and
resemble the charge-neutral conformation shown in Figure 3c. As in
ferroelastics, there can be large stresses associated with such triple
domain junctions\cite{Salje90,Pertsev92} However, such stresses are
qualitatively similar for all tip structures, regardless of domain
orientation with respect to the applied electric field, and can not
explain the difference in their field-induced mobility. We expect that
the domain behavior we observe is characteristic of a ferroelectric
system that is not mechanically clamped. Examples would be
polycrystalline films, and films with low Young's modulus contact
electrodes, such as liquids and some conducting oxides.

\section{Conclusions}

We have presented evidence, using in-situ transmission microscopy on
freestanding \knb\ and \bat\ samples, that in a polyaxial ferroelectric
not all domains of a given polarization orientation are equally
switchable in low applied fields.  Angled and curved domain walls are
abundant, and often occur when perpendicular sets of \dwn\ intersect.
We can understand the observed domain configurations as resulting from a
competition between the minimization of the electrostatic energy and
conservation of the polarization charge.  In our experiments, domain
switching occurs as a fluid, non-rigid, wall motion and all observed
domain patterns are stable to mechanical tapping. Spontaneous domain
creation is never observed at fields well below the coercive field.  The
switching efficiency of a particular domain is determined by its allowed
final states which are defined by its neighbors.  If the relative
energetics are unfavorable, switching will be inhibited.  Such domain
interlocking results in a reduction of the switchable volume in the
sample, and provides an inherent contribution to degradation effects in
polyaxial ferroelectrics.  The sequential depinning of domain needle
tips that we observe in \bat\ and \knb\ under applied electric fields,
reveals that polarization charge must be flowing along domain walls that
are experiencing an electrostatic switching pressure.  Polarization
charge ripples facilitate domain switching in a manner roughly analogous
to the role played by dislocations in crystal slip.

Mobile needle-like domains retract in applied fields because of the
reduced switching barriers at the tips.  Inhibited domains have an
environment that energetically prohibits the formation of a charged tip
and thus require indirect switching processes.  Rotation of the
direction of the applied electric field through $90^{\circ}$ should free
such inhibited domains.  In addition, complex interactions with mobile
extrinsic charges can further inhibit the retraction of intrinsically
charged needles.  Such processes could contribute to a reduction in the
dipolar screening charge flowing in the external circuit to the
electrodes, and thus could contribute to ferroelectric fatigue and
possibly imprint.

Domain interlocking is less likely to occur in ferroelectrics with
reduced degrees of polarization freedom.  We expect that ferroelectric
fatigue will be greatly diminished in materials with a single preferred
polarization axis, possibly induced by application of stress or field.

We thank S.\ Bhattacharya, J.\ D.\ Chadi and J.\ F.\ Scott for
stimulating discussions.

\newpage
\section{Appendix}
\begin{center}\textbf{Impact of wall tilt on switching barrier and
switching pressure.}\end{center}

We examine the simple isotropic Landau-Ginsburg free energy of a
freestanding ferroelectric region close to a \dwo\ domain wall that is
tilted from the charge-neutral orientation by an angle $\theta$.

Figure 7a shows a model ferroelectric capacitor with a \dwo\ domain wall
perpendicular to the electrodes.  The wall has been sheared slightly
such that, near the middle of the wall, a small region has tilted
through an angle $\theta$.  Far from the wall, the polarizations in each
domain are $\pm\mbox{\boldmath $P$}_{0}$, parallel and antiparallel to
the electrode normals. A surface polarization charge develops at the
tilted wall region. However, the charge at the wall is not simply
$\sigma = 2P_{0}\sin\theta$. There is a depolarization field
$\mbox{\boldmath $E$}_{\rm w} = E_{\rm w}\hat{\mbox{\boldmath $n$}}$
associated with the wall charge, that depolarizes the local
polarizations to $\mbox{\boldmath $P$}_{1} = \pm(\mbox{\boldmath
$P$}_{0} + \Delta \mbox{\boldmath $P$})$. Assigning
$\hat{\mbox{\boldmath $n$}}$ to be the unit vector along the tilted wall
normal, the self-consistent surface charge $\sigma_{\rm w}$ at the wall
is,
\begin{eqnarray}
         \sigma_{\rm w} & = & \left[ \left( -\mbox{\boldmath $P$}_{0} +
         \Delta\mbox{\boldmath $P$} \right) - \left( +\mbox{\boldmath
         $P$}_{0} + \Delta\mbox{\boldmath $P$} \right) \right] \!\cdot\!
         \hat{\mbox{\boldmath $n$}} \nonumber \\ & = & -2\mbox{\boldmath
         $P$}_{0} \!\cdot\!  \hat{\mbox{\boldmath $n$}} -
         2\epsilon_{0}\chi E_{\rm w} \hat{\mbox{\boldmath $n$}} \!\cdot\!
         \hat{\mbox{\boldmath $n$}} \nonumber \\ & = & 2P_{0}\sin\theta -
         2\epsilon_{0}\chi E_{\rm w}.
      \label{eq:sigma180}
\end{eqnarray}

The electric field $E_{\rm w}$ is also related to the surface charge
$\sigma_{\rm w}$ by
\begin{equation}
      E_{\rm w} = \frac{\sigma_{\rm w}}{2\epsilon\epsilon_{0}}.
      \label{eq:Ew}
\end{equation}
The field is of opposite sign in each domain, and can be considered
constant out to a distance equal to the wall length, i.e.\ within the
cylindrical region shown in Figure 7b.

Substituting for $E_{\rm w}$ into equation~(\ref{eq:sigma180}) gives
\begin{equation}
      \sigma_{\rm w} = \frac{2P_{0}\sin\theta}{1 + \chi/\epsilon}
      \approx P_{0}\sin\theta
      \label{eq:sigma180_2}
\end{equation}
and
\begin{equation}
      E_{\rm w} = \frac{P_{0}\sin\theta}{\epsilon\epsilon_{0}
      \left(1 + \chi/\epsilon \right)} \approx
      \frac{P_{0}\sin\theta}{2\epsilon\epsilon_{0}}.
      \label{eq:Ew2}
\end{equation}

Thus, in the limit of large $\epsilon$, the depolarization field $E_{\rm
w}$ acts to reduce the wall charge to about half of its nominal value,
$2P_{0}\sin\theta$.  The depolarization field results in the local
polarization vectors being tilted away from the wall by an amount
$\theta_{1}$.  $\theta_{1}$ is found from the triangle defined by
$\mbox{\boldmath $P$}_{0}$, $\mbox{\boldmath $P$}_{1}$ and $\Delta
\mbox{\boldmath $P$}$ in Figure 7b, from which we get the sine rule
relations

\begin{equation}
      \frac{P_{0}}{\cos(\theta \!-\!\theta_{1})} =
      \frac{P_{1}}{\cos\theta} =
      \frac{\chi P_{1}\sin(\theta \!-\!\theta_{1})}{\epsilon \sin\theta_{1}}
      \label{eq:sinerule}
\end{equation}
which give
\begin{equation}
      \cos\theta\sin(\theta \!-\!  \theta_{1}) -
      \frac{\epsilon}{\chi}\sin\theta_{1} = 0.
      \label{eq:sintheta}
\end{equation}

There is an additional volume charge that appears away from the domain
wall.  Far from the tilted wall, the polarization returns to the bulk
value $\mbox{\boldmath $P$}_{0}$.  The associated $\mbox{\boldmath
$\nabla$} \!\cdot\!\mbox{\boldmath $P$}$ gives rise to a volume
polarization charge.  The total volume charge is equivalent to the
charge ``missing'' from the wall due to depolarization, thus the total
wall plus volume charge equals $2P_{0}\sin\theta$ times the tilted wall
area.

An additional complication is imposed by the equipotential boundary
conditions at the electrodes.  These boundary conditions are equivalent
to creating mirror images of the charged region.  We have assumed that
the multipole contributions to the local field can be ignored.

The surface charge density on a wall of fixed tilt is independent of the
electric field, and equal to $\sigma = 2\mbox{\boldmath $P$}_{1}
\sin(\theta \!-\! \theta_{1})$. This can be rewritten in terms of the
local polarization {\boldmath $P$} as $\sigma = 2|\mbox{\boldmath $P$}
\!-\! \epsilon_{0} \chi \mbox{\boldmath $E$}| \sin(\theta \!-\!
\theta_{1})$ (see Figure 7b).  ({\boldmath $P$} = {\boldmath $P$}$_{\rm
L}$ or {\boldmath $P$}$_{\rm R}$ depending on which domain we are
considering.)  Incorporating the depolarization energy, $U_{\rm W} =
\frac{1}{2} \epsilon \epsilon_0 E^2_{\rm w}$ into a standard
Landau-Ginsburg free energy expansion, we obtain for the regions either
side of a tilted \dwo\ domain wall
\begin{equation}
         F \approx \frac{1}{2}\alpha\left[T - \tilde{T}_{\rm
         c}(\theta)\right] P^{2} + \frac{\beta}{4}P^{4} +
         \frac{1}{6}\gamma P^{6} - \left[1 +
         g(\theta)\right]\mbox{\boldmath $P$} \!\cdot\!  \mbox{\boldmath
         $E$} + \frac{1}{2}\mu \xi^{2} \left(\mbox{\boldmath
         $\nabla$} \!\cdot\! \mbox{\boldmath $P$} \right)^{2}
         \label{eq:F3}
\end{equation}
where the reduced effective Curie temperature $\tilde{T}_{\rm
c}(\theta)$ is given by $\tilde{T}_{\rm c}(\theta) = T_{\rm c} \!-\!
g(\theta)/\epsilon\epsilon_{0}\alpha$, with $g(\theta) = \sin^{2}(\theta
\!-\! \theta_{1})$.  Since the regions of interest are just either side
of the domain wall, at a distance much closer than the characteristic
length scale $\xi$ for depolarization field gradients, the last term is
essentially negligible.

 From this simple isotropic analysis, it emerges that one effect of wall
tilt on a \dwo\ domain is to reduce the barrier to switching by
decreasing the effective Curie temperature $\tilde{T}_{\rm c}(\theta)$
of domain regions close to the tilted wall. In addition, the term
corresponding to the electrostatic pressure on the wall is enhanced by a
factor $\left[1 + g(\theta)\right]$. This enhancement has the effect of
maintaining an approximately constant electrostatic pressure
$2\mbox{\boldmath $P$}_{0} \!\cdot\! \mbox{\boldmath $E$}$ on the tilted
wall, despite the fact that the local polarization {\boldmath $P$}$_{1}$
is reduced by the depolarization field.

The expression for the effective Curie temperature $\tilde{T}_{\rm c} =
T_{\rm c} \!-\!  \eta g(\theta)$, where $\eta =
1/\epsilon\epsilon_{0}\alpha$ ($\equiv 4\pi/\epsilon\alpha$ c.g.s.),
that emerges from equation~(\ref{eq:F3}) indicates that there exists a
threshold angle, defined by $g(\bar{\theta}_{\rm T}) = (T_{\rm c} \!-\!
T)/\eta$, at each temperature delineating the ferroelectric and
paraelectric phases. The presence of $\bar{\theta}_{\rm T}$ is
consistent with our TEM studies, which show typical domain tip angles in
\bat\ of $\theta_{\rm B}^{*} = 5 \pm 1.4^{\circ}$ and in \knb\ of
$\theta_{\rm K}^{*} \le 7^{\circ}$. Noting that $\eta = C/\epsilon$,
where $C$ is the Curie constant, we find using standard values
\cite{Jona93} for $C$ and $\epsilon$ for \bat\ and \knb, that $\eta_{\rm
K} < \eta_{\rm B}$; thus the observation that $\theta_{\rm K}^{*} >
\theta_{\rm B}^{*}$ is consistent with our expectations from this
phenomenological approach \cite{Scott01}.

The coercive field, $E_{\rm c}(\theta)$, required to move a tilted \dwo\
domain wall varies with $\theta$ as
\begin{equation}
        E_{\rm c}(\theta) = \frac{E_{\rm c}(0)}{\left[1 +
        g(\theta)\right]} \left[1 - \frac{\eta g(\theta)}{T_{\rm c} - T
        - \eta g(\theta)}\right]^{3/2}.
        \label{eq:Ec}
\end{equation}
Since $0 \!\le\!  g(\theta) \!\le\!  1$, $g(\theta)$ acts to reduce
$E_{\rm c}(\theta)$ locally.  Again, there exists a threshold angle at
each temperature, $\bar{\theta}_{E} < \bar{\theta}_T$ defined by
$g(\bar{\theta}_{E}) = (T_{\rm c} - T)/2 \eta$, above which there is
field-induced switching.  This angular dependence could provide a major
contribution to the spread in coercive fields observed in standard
hysteresis loops.

A more complete free energy expression would incorporate anisotropy in
both the polarization and stress fields.  Such refinements are important
for modeling the detailed behavior and shape of domain needles in
different materials. Strain energy is particularly important for
materials supported on stiff substrates where domain clamping effects
dominate. Isotropic elastic strain fields are long-range and, unlike the
electric fields, can not be screened. They are expected to increase
switching barriers overall \cite{Littlewood86}, but should not affect
the angular dependencies discussed here. Clearly, they will not affect
the electrostatic pressure.  In our experiments, the ferroelectrics are
freestanding and are not clamped by any support material. Further, the
initial tip structures are similar for both mobile and immobile needles
(i.e. Domains A and C versus domain B in figure 1). Elastic constraints
caused by tip stresses are expected to be similar for all these domains,
and therefore can not explain the different domain behaviors in the
isotropic limit. Thus the simplified isotropic treatment presented here
is adequate for understanding the underlying physical insights revealed
by our data.

\begin{figure}[ht!]
	\centerline{\epsfig{file=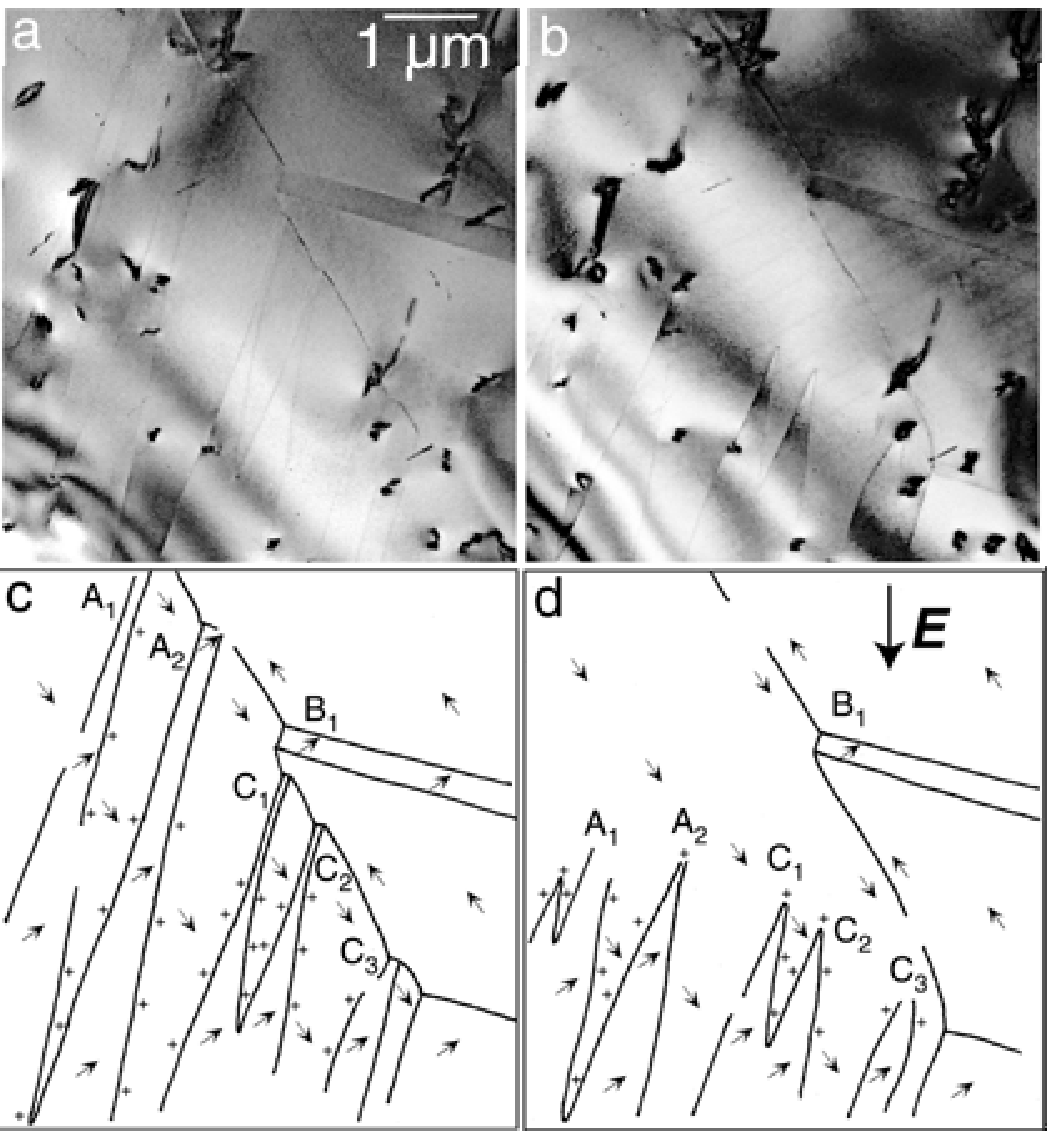}} \vspace{0.75in}
	\caption{Bright-field in-situ transmission electron micrographs
	of thin \knb\ (a) before and (b) after application of an
	electric field, with accompanying schematics (c) and (d)
	indicating the electric field direction, polarization directions
	and accompanying boundary charges.}
\label{fig1}
\end{figure}
\newpage

\begin{figure}[ht!]
        \centerline{\epsfig{file=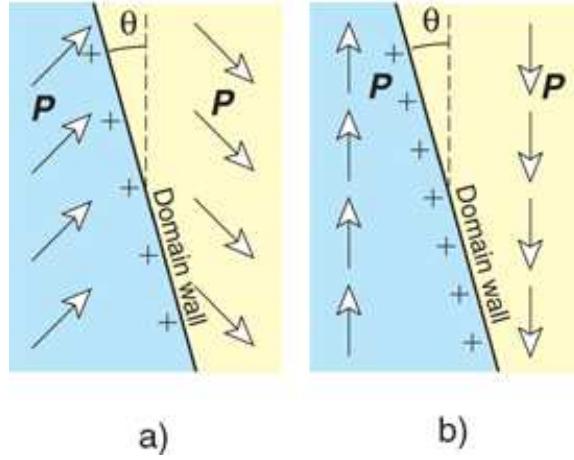,width=3in}}
        \vspace{10pt}
        \caption{Geometry of tilted; a) \dwn domains; b) \dwo domains.}
\label{fig2}
\end{figure}

\begin{figure}[hb!]
       \vspace{1in} \centerline{\epsfig{file=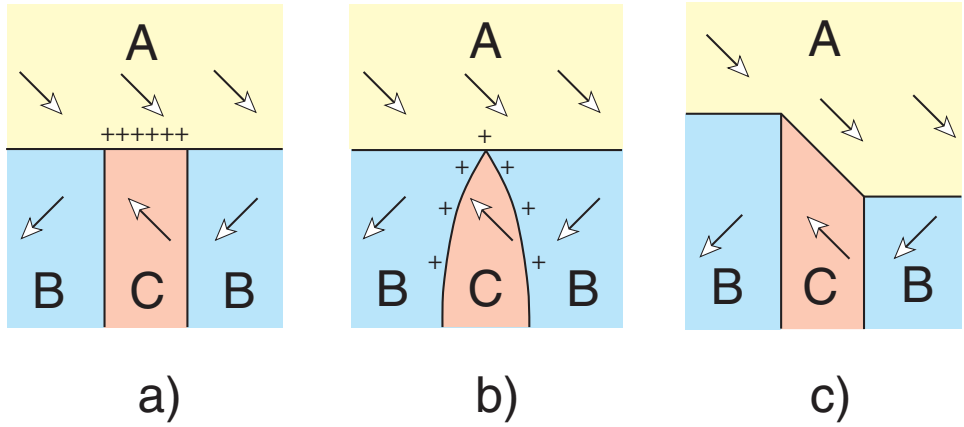,width=5in}}
       \vspace{10pt} \caption{Diagram illustrating how polarization
       charge can occur when 3 or more perpendicular \dwn\ domain
       orientations meet. a) All domain walls are straight.  polarization
       charge $Q$ is localized to the \dwfv\-tilted \dwo\ wall between
       domains A and C. b) Curved walls on domain C. The same
       polarization charge $Q$ is now spread over the walls between B and
       C. c) Domains A and B move to tilt the wall between A and C
       transforming it into an untilted \dwo\ domain wall.  Polarization
       charge is eliminated.}
\label{fig3}
\end{figure}
\newpage

\begin{figure}[ht!]
       \centerline{\epsfig{file=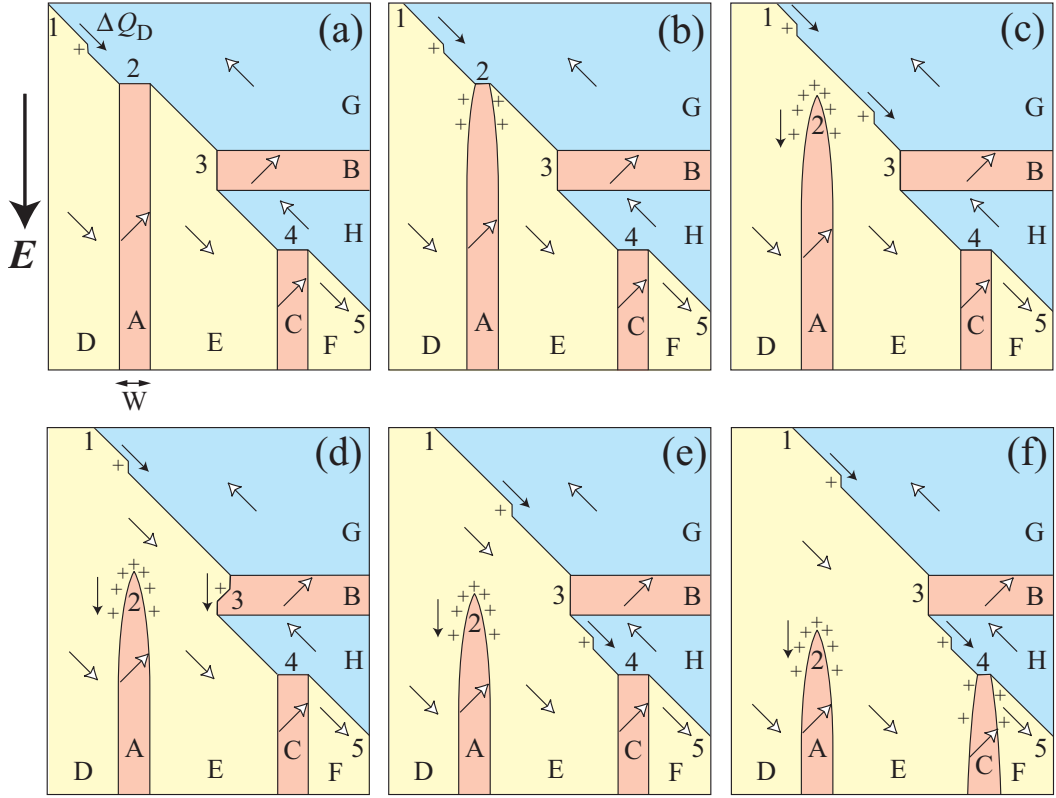,width=5.5in}} \vspace{0.25in}
       \caption{Schematic of domains observed in Figure 1.  On
       application of a vertical electric field, positive polarization
       charge flows along the diagonal wall until it reaches domain $A$
       (a).  Polarization charge is pinned by the tip of $A$ resulting in
       domain curvature at the tip until the tip depins, (b) and (c).
       The needle domain $A$ retracts under the field.  Domain $B$ cannot
       support positive polarization charge because it is energetically
       unfavorable (d).  The polarization charge bypasses domain $B$
       where it is subsequently sequestered by domain $C$, (e) and (f).
       Domain C detaches after domain A. Although domain $B$ has the
       polarization orientation as domains $A$ and $C$, its switching is
       inhibited by its environment.}
\label{fig4}
\end{figure}
\newpage

\begin{figure}[ht!]
        \centerline{\epsfig{file=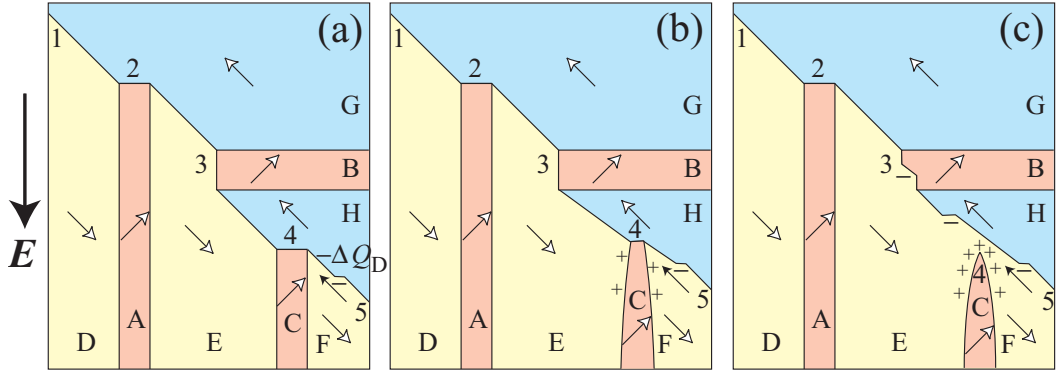,width=5.5in}} \vspace{0.25in}
        \caption{The domain pattern of Figure 4a, with the same field
        direction, but with negative charge ripples flowing upwards (a).
        (b) The charge arrives at domain C first after converting part of
        domain H into domain F by moving the wall region 4--5 upwards.
        The tip region 4 of domain C develops a positive charge because
        part of domain C as well as part of domain H is being switched
        into domains E and F. Domain C acts as a sink of charge until
        domain C detaches, (c) whereupon charge continues towards region
        3.  As before, the tip of domain B does not form a point.  The
        principal difference with Figure 4 is that here domain C detaches
        before domain A.}
\label{fig5}
\end{figure}
\newpage

\begin{figure}[ht!]
         \centerline{\epsfig{file=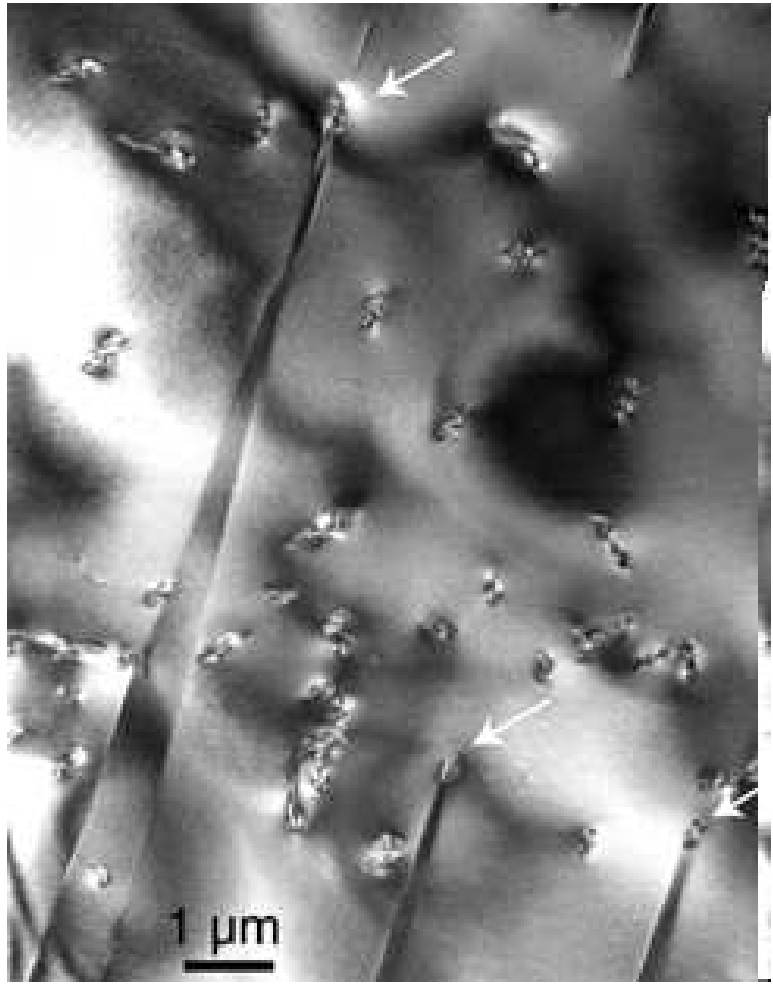,width=4in}} \vspace{0.75in}
         \caption{Bright-field in-situ transmission electron micrograph
         of \knb\ showing three $90^\circ$ needle-like domains pinned at
         their tips by dislocations.}
\label{fig6}
\end{figure}

\begin{figure}[hb!]
      \vspace{1in} \centerline{\epsfig{file=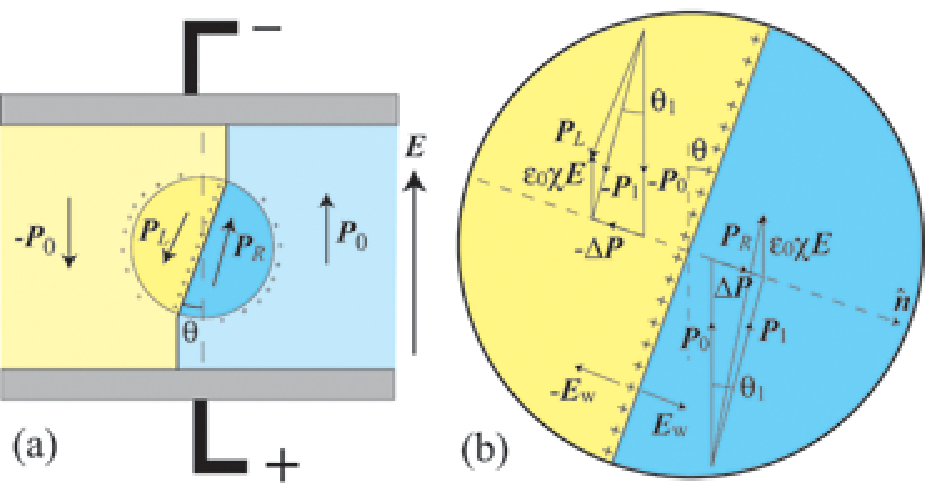,width=5in}}
       \vspace{10pt} \caption{(a) Diagram illustrating the local
       polarization charges and polarization vectors $\mbox{\boldmath
       $P$}_{\rm L}$ and $\mbox{\boldmath $P$}_{\rm R}$ either side of a
       tilted \dwo\ domain wall far from the electrodes.  (b) Detailed
       view of the polarization geometry near the tilted wall region.  In
       zero appled field, the local polarization is tilted $\theta_{1}$
       away from the wall.  The depolarization field $E_{\rm w}$ is
       treated as constant out to a distance equal to half the lateral
       extent of the tilted wall region.}
\label{fig7}
\end{figure}
\newpage


\begin{references}

\bibitem{Scott00}
J.F. Scott, {\sl Ferroelectric Memories} (Springer-Verlag, Berlin, 2000).

\bibitem{Colla98a}
E.L. Colla, S. Hong, D.V. Taylor, A.K. Tagantsev, N. Setter and K. No,
{\sl Appl. Phys. Lett.} {\bf 72}. 2763 -- 2765 (1998).

\bibitem{Ramesh93}
R. Ramesh, H. Gilchrist, T. Sands, V.G. Keramidas, R. Haakenaasen and
D.K. Fork,
{\sl Appl. Phys. Lett.} {\bf 63}, 3592 -- 94 (1993).

\bibitem{Nakamura94}
T. Nakamura, Y. Nakao, A. Kamisawa and H. Takasu,
{\sl Appl. Phys. Lett.}
{\bf 65}, 1522 (1994).

\bibitem{DeAraujo95}
C.A. Paz de Araujo, J.D. Cuchiaro, L.D. McMillan, M.C. Scott and
J.F. Scott,
{\sl Nature} {\bf 374}, 627 -- 629 (1995).

\bibitem{Scott98}
J.F. Scott,
{\sl Ferroelectrics Review}, {\bf 1}, 129
(1998) and references therein.

\bibitem{Damjanovic98}
D. Damjanovic,
{\sl Rep. Prog. Phys.} {\bf 61}, 1267 -- 1324 (1998).

\bibitem{Colla98}
E.L. Colla, D.V. Taylor, A.K. Tagantsev and N. Setter,
{\sl Appl. Phys. Lett.} {\bf 72}. 2478 -- 2480 (1998).

\bibitem{Warren95a}
W.L. Warren, D. Dimos, B.A. Tuttle, G.E. Pike, R.W. Schwartz,
P.J. Clews
and D. McIntyre,
{\sl J. Appl. Phys.} {\bf 77}, 6695 -- 6702 (1995).

\bibitem{Scott89}
J.F. Scott and C.A. Paz de Araujo,
{\sl Science} {\bf 246}, 1400
(1989).

\bibitem{Tanaka64}
M. Tanaka and H. Goro,
{\sl Journal of the Physical
Society of Japan} {\bf 19}, 954, 954--970 (1964).

\bibitem{Hu86}
Y.H. Hu, H.M. Chan, X.Z. Wen and M.P Warmer,
{\sl Journal of
the American Ceramic Society}, {\bf 69}, 594 -- 602 (1986).

\bibitem{Yamamoto80}
N. Yamamoto, K. Yagi and G. Honjo,
{\sl Physica Status
Solidi (a)} {\bf 62}, 657 (1980).

\bibitem{Snoeck94}
E. Snoeck, L. Normand, A. Thorel and C. Roucau,
{\sl Phase Transitions} {\bf 46}, 77--88 (1994).

\bibitem{Krishnan98}
A. Krishnan, M.E. Bisher and M.M.J. Treacy,
{\sl Mat. Res. Soc. Symp. Proc.}, {\bf 541}, 475 (1998).

\bibitem{Kilaas88}
R. Kilaas, {\sl Macintosh Program for Simulation of High Resolution
TEM Images}, (Lawrence Berkeley Laboratories, Berkeley, 1988)..

\bibitem{Salje90}
E.K.H. Salje,
{\sl Phase Transitions in Ferroelastic and Co-Elastic Crystals}
(Cambridge University Press, Cambridge, 1990).

\bibitem{Nishiyama78}
Z. Nishiyama,
{\sl Martensitic Transformation}
(Academic Press, New York, 1978).

\bibitem{Chervonobrodov88}
S. P. Chervonobrodov and A. L. Roytburd,
{\sl Ferroelectrics} {\bf 83}, 109 (1988).

\bibitem{Callaby65}
D.R. Callaby,
{\sl J. Appl. Phys.} {\bf 36} 2761 (1965).

\bibitem{Jona93}
F. Jona and G. Shirane, {\sl Ferroelectric Crystals},
(Dover, New York, 1993).

\bibitem{Landauer57}
R. Landauer, {\sl J. Appl. Phys.} {\bf 28}, 227 (1957).

\bibitem{Miller58}
R.C. Miller and A. Savage, {\sl Phys. Rev.} {\bf 112}, 755 (1958).

\bibitem{Krishnan00}
{\sl Fundamental Physics of Ferroelectrics 2000}, ed. R.E. Cohen
(AIP Conference Proceedings, New York, 2000).

\bibitem{Scott88}
J. F. Scott and B. Pouligny,
{\sl J. Appl. Phys.} {\bf 64}, 1547 (1988).

\bibitem{Bornand00}
V. Bornand, S. Trolier-McKinstry, K. Takemura and C.A. Randall,
{\sl J. of Applied Phys.} {\bf 87}, 3965 (2000).

\bibitem{Pan92}
W.Y. Pan, C.F. Yue and B.A. Tuttle,
{\sl Ceram. Trans.} {\bf 25}, 385--397 (1992).

\bibitem{Robels95}
U. Robels, H.J. Calderwood and G. Arlt, {\sl J. Appl. Phys.}
{\bf 77}, 4002 (1995);
G. Arlt and H. Neumann, {\sl Ferroelectrics} {\bf 87}, 109 (1988).

\bibitem{Yoo92}
I.K. Yoo and S.B. Desu, {\sl Phys. Status Solidi A}
{\bf 133}, 565 (1992).

\bibitem{Brennan93}
C. Brennan, {\sl Ferroelectrics} {\bf 150}, 199 (1993).

\bibitem{Dawber00}
M. Dawber and J.F. Scott,
{\sl Appl. Phys. Lett.} {\bf 76} 1060--1062 (2000).

\bibitem{Pertsev88} N.A. Pertsev,
{\sl Sov. Phys. Solid State} {\bf 30}, 1616-1621 (1988).




\bibitem{Ganpule01}
It has been observed that \dwn\ domain walls do not move in clamped
single crystal PZT films under applied electric fields. However, they
do move once freed from the mechanical constraints imposed by the
substrate (C.\ Ganpule and R.\ Ramesh, private communication, 2001).

\bibitem{Pertsev92}
N. Pertsev and G. Arlt, {\sl Ferroelectrics} {\bf 132} 27-40 (1992).

\bibitem{Scott01}
We thank J.\ F.\ Scott for alerting this to this point.

\bibitem{Littlewood86}
P.B. Littlewood and P. Chandra,
{\sl Phys. Rev. Lett.} {\bf 57}, 2415 (1986);
P. Chandra,
{\sl Phys. Rev. A} {\bf 39}, 3672 (1989).

\end{references}
\end{document}